\documentclass[aps,prb,twocolumn,superscriptaddress,showpacs]{revtex4}

\usepackage{graphicx}
\usepackage{dcolumn}
\usepackage{bm}

\begin{document}

\title{Low-temperature heat transport of Nd$_2$CuO$_4$: Roles of Nd magnons
and spin-structure transitions}

\author{Z. Y. Zhao}
\affiliation{Hefei National Laboratory for Physical Sciences at
Microscale, University of Science and Technology of China, Hefei,
Anhui 230026, People's Republic of China}

\author{X. M. Wang}
\affiliation{Hefei National Laboratory for Physical Sciences at
Microscale, University of Science and Technology of China, Hefei,
Anhui 230026, People's Republic of China}

\author{B. Ni}
\affiliation{Hefei National Laboratory for Physical Sciences at
Microscale, University of Science and Technology of China, Hefei,
Anhui 230026, People's Republic of China}

\author{Q. J. Li}
\affiliation{Hefei National Laboratory for Physical Sciences at
Microscale, University of Science and Technology of China, Hefei,
Anhui 230026, People's Republic of China}

\author{C. Fan}
\affiliation{Hefei National Laboratory for Physical Sciences at
Microscale, University of Science and Technology of China, Hefei,
Anhui 230026, People's Republic of China}

\author{W. P. Ke}
\affiliation{Hefei National Laboratory for Physical Sciences at
Microscale, University of Science and Technology of China, Hefei,
Anhui 230026, People's Republic of China}

\author{W. Tao}
\affiliation{Hefei National Laboratory for Physical Sciences at
Microscale, University of Science and Technology of China, Hefei,
Anhui 230026, People's Republic of China}

\author{L. M. Chen}
\affiliation{Department of Physics, University of Science and
Technology of China, Hefei, Anhui 230026, People's Republic of
China}

\author{X. Zhao}
\affiliation{School of Physical Sciences, University of Science
and Technology of China, Hefei, Anhui 230026, People's Republic of
China}

\author{X. F. Sun}
\email{xfsun@ustc.edu.cn}

\affiliation{Hefei National Laboratory for Physical Sciences at
Microscale, University of Science and Technology of China, Hefei,
Anhui 230026, People's Republic of China}

\date{\today}

\begin{abstract}

We report the magnetic-field dependence of thermal conductivity
($\kappa$) of an insulating cuprate Nd$_2$CuO$_4$ at very low
temperatures down to 0.3 K. It is found that apart from the
paramagnetic moments scattering on phonons, the Nd$^{3+}$ magnons
can act as either heat carriers or phonon scatterers, which
strongly depends on the long-range antiferromagnetic transition
and the field-induced transitions of spin structure. In
particular, the Nd$^{3+}$ magnons can effectively transport heat
in the spin-flopped state of the Nd$^{3+}$ sublattice. However,
both the magnon transport and the magnetic scattering are quenched
at very high fields. The spin re-orientations under the in-plane
field can be conjectured from the detailed field dependence of
$\kappa$.

\end{abstract}


\maketitle

\section{Introduction}

The insulating parent compounds of cuprate superconductors are
known to have an antiferromagnetic (AF) order of Cu$^{2+}$ spins.
Because of the quasi-two-dimensionality of the Cu$^{2+}$ spin
structure, these materials can show rather strong magnon heat
transport at relatively high temperatures.\cite{Sun_LCO, Hess_LCO,
Sun_PLCCO, Berggold, Sun_PLCO, Jin_NCO} However, the magnon
transport seems to be negligible at low temperatures, probably due
to the anisotropy gap of the spin spectrum. The external magnetic
field can strongly suppress or close the gap at some critical
fields, usually accompanied with the spin-flop transitions. It was
therefore expected to observe the magnon heat transport in
magnetic field.\cite{Jin_NCO} In last several years, a few works
on this topic have been performed on parent compounds of
electron-doped cuprate superconductors,
Pr$_{1.3}$La$_{0.7}$CuO$_4$ (PLCO) and Nd$_2$CuO$_4$
(NCO).\cite{Sun_PLCO, Jin_NCO, Li_NCO} The low-$T$ heat transport
of PLCO seemd to be quite simple and indicated that some
paramagnetic moments such as ``free" Cu$^{2+}$ spins can scatter
phonons and lead to non-negligible magnetic-field dependence of
phononic thermal conductivity.\cite{Sun_PLCO} It is notable that
the PLCO data did not indicate the possibility of magnon heat
transport, although the spin-flop transition can happen for not
strong in-plane fields. On the other hand, the NCO results are
more complicated because of the complexity of magnetism caused by
the Nd$^{3+}$ ions.\cite{Jin_NCO, Li_NCO}

It is known that in NCO the Cu$^{2+}$ spins order
antiferromagnetically below $T_N = 245\sim255$ K with a
noncollinear magnetic structure,\cite{Matsuda, Endoh, Cherny,
Structure, Structure2, Richard, NCO_neutron} that is, the Cu spins
in the same CuO$_2$ plane are antiferromagnetically ordered and
those in the adjacent planes along the $c$ axis are perpendicular
to each other. The moment of Nd$^{3+}$ ion is enhanced drastically
with lowering temperature and a long-range AF order develops below
$\sim$ 1.5 K with the same noncollinear spin structure as
Cu$^{2+}$ spins.\cite{Structure, NCO_neutron, Lynn} When the
magnetic field is applied in the CuO$_2$ plane, the Cu$^{2+}$
spins can re-orientate and enter a spin-flopped
state.\cite{Structure, Structure2, Cherny, Thalmeier, NCO_SF1,
NCO_SF2,NCO_SF3, NCO_SF4} The transition fields are reported to be
4.5 T and 0.75 T for $H \parallel a$ and $H \parallel [110]$,
respectively.\cite{Cherny, NCO_SF3} In addition, the strong
coupling between Nd$^{3+}$ spins and Cu$^{2+}$ spins, which was
found to be about 4 T,\cite{Richard} drives Nd$^{3+}$ spins to
rotate together with Cu$^{2+}$ spins, keeping their relative
orientation unchanged, and forms a three-layer unit consisting of
one Cu$^{2+}$ layer sandwiched between two Nd$^{3+}$ layers as a
whole. For this reason, the Nd$^{3+}$ spins are generally
considered to change together with Cu$^{2+}$ sublattice under the
influence of magnetic field. However, at low temperatures ($<$ 1.5
K) when the AF order of Nd$^{3+}$ spins is formed, the magnetic
structures and field-induced transitions of Nd$^{3+}$ sublattice
are still unclear since all the earlier neutron scattering in
applied field were carried out at relatively high temperatures.

Two earlier studies on the low-$T$ thermal conductivity of NCO
have revealed strong magnetic-field dependence of $\kappa$, which
was believed to be associated with the spin flop of Cu$^{2+}$
spins.\cite{Jin_NCO, Li_NCO} An additional conduction channel of
Nd$^{3+}$ magnons was supposed to be the reason for the
enhancement of $\kappa$ in high fields, considering the
reorientation of Nd$^{3+}$ spins under Nd$^{3+}$-Cu$^{2+}$
interaction. However, there are some quite unclear issues calling
for more careful investigations. First, it is apparently
questionable in Ref. \onlinecite{Jin_NCO} to attribute the large
increase of $\kappa$ above some transition fields to the Nd$^{3+}$
magnons at temperatures above 2 K, where the long-range ordering
of Nd$^{3+}$ spins has been lost. In particular, the field-induced
enhancement of $\kappa$ is even much larger at 5 K than that at 2
K.\cite{Jin_NCO} In that work, more direct evidence for Nd$^{3+}$
magnon transport, which should be detected below the N\'eel
transition of Nd$^{3+}$ spins, had not been investigated. Another
work performing measurements down to milli-Kelvin temperatures,
however, did not report the magnetic-field dependent
data,\cite{Li_NCO} which is indispensable for showing the
relationship between heat transport and the transition of magnetic
structure. Second, although the low-energy magnons can be
significantly excited at the spin-flop transition and may
contribute to carrying heat, because of the closure of the gap in
magnon spectra, further increasing magnetic field will always
shift the magnon dispersion upward and finally suppress the
low-energy magnon excitations.\cite{Spin_flop} So it is odd to
assume that the magnon heat transport induced by the spin flop
transition can be still active in the high-field
limit.\cite{Jin_NCO, Li_NCO} Third, it was assumed that the
Nd$^{3+}$ spin lattice should rotate together with Cu$^{2+}$
sublattice in the in-plane field. However, it is notable that the
Nd$^{3+}$-Nd$^{3+}$ interaction can be enhanced significantly at
very low temperatures due to the increased magnitude of Nd$^{3+}$
moment, which means that the Nd$^{3+}$ sublattice could
re-orientate independently. In this work, we study the temperature
and magnetic-field dependencies of $\kappa$ of high-quality NCO
single crystals in great detail to investigate the role of magnons
in the low-$T$ heat transport. Our results on $\kappa(H)$
isotherms demonstrate that the phonon conductivity is strongly
dependent on the magnetic field, and the Nd$^{3+}$ magnon can play
a dual role in the transport as the heat carriers or the phonon
scatterers. To be exact, the Nd$^{3+}$ magnons mainly scatter
phonons at the critical region (1$\sim$2 K) of N\'eel transition
in zero field, while they can significantly transport heat in the
spin-flopped state induced by the in-plane field. Intriguingly,
the drastic behaviors of $\kappa(H)$ under the in-plane field
reveal the spin re-orientation transitions of Nd$^{3+}$ sublattice
at sub-Kelvin temperatures.

\section{Experiments}

High-quality Nd$_2$CuO$_4$ single crystals are grown by using the
slow cooling method with CuO$_2$ as a self flux. The plate-like
samples for transport measurements are confirmed to grow along the
$ab$-plane by the x-ray back-reflection Laue photographs. The
dimensions along the $c$ axis of NCO crystals are so small that
the measurement with heat current along the $c$ axis is difficult
to carry out. The thermal conductivity is therefore measured only
along the $a$ axis by using a conventional steady-state technique
and two different processes: (i) using a ``one heater, two
thermometers" technique in a $^3$He refrigerator and a 14 T magnet
at temperature region of 0.3 -- 8 K; (ii) using a
Chromel-Constantan thermocouple in a pulse-tube refrigerator for
the zero-field data above 4 K.\cite{Wang_HMO, Zhao_GFO, Sun_DTN}

\section{Results and Discussion}

\begin{figure}
\includegraphics[clip,width=7cm]{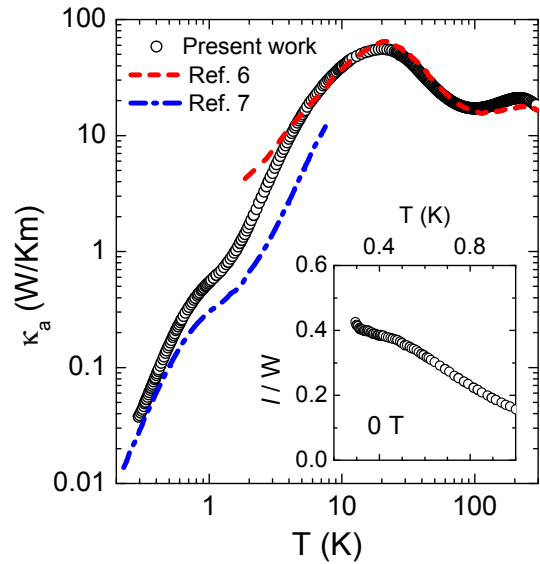}
\caption{(Color online) The $a$-axis thermal conductivity of
Nd$_2$CuO$_4$ single crystal in the zero field, compared with the
results reported by Jin {\it et al.}\cite{Jin_NCO} and Li {\it et
al.}\cite{Li_NCO} Inset: the ratio of the phonon mean free path
$l$ to the averaged sample width $W$ in the zero field.}
\end{figure}

Figure 1 shows the temperature dependence of thermal conductivity
of a NCO single crystal in zero field, together shown are data
from some other groups.\cite{Jin_NCO, Li_NCO} The large peak at 20
K is apparently the so-called phonon peak in
insulators.\cite{Berman} It is known that the magnitude of the
phonon peak is dominated by the phonon scattering by the crystal
defects and impurities and is therefore a good characterization
for the crystal quality.\cite{Berman} It is clear that the phonon
peak of our crystal is almost the same as the data obtained on a
NCO crystal grown using the floating-zone method,\cite{Jin_NCO}
suggesting the high quality of our NCO crystal. Although the
high-quality NCO had not been measured down to very low
temperatures in the earlier work,\cite{Jin_NCO} our crystal can
alternatively demonstrate the intrinsic low-$T$ heat transport of
this compound. Below 10 K, the thermal conductivity decreases
quickly with lowering temperature, like usual insulators, but
there is an obvious variation in slope of the $\kappa(T)$ curve
around 1.5 K, which is apparently related to the AF ordering of
Nd$^{3+}$ ions.\cite{Structure, NCO_neutron, Lynn} However, only
from the zero-field data, it is not clear whether the slope change
is caused by an enhanced magnon scattering on phonons at the
critical region of AF phase transition or an appearance of the
magnon heat transport below the transition. The magnetic-field
dependence of $\kappa$ is known to be useful to clarify this
issue. Before analyzing the $\kappa(H)$ data, we can make an
estimation of the phonon mean free path ($l$) at very low
temperatures assuming that the magnons are not acting as heat
carriers. Following a standard calculation,\cite{Sun_Comment} $l$
can be obtained from the kinetic formula
$\kappa=\frac{1}{3}C\bar{v}\l$, where $C = \beta T^3$ is the
phonon specific heat and $\bar{v}$ is the averaged sound velocity,
and both $\beta$ and $\bar{v}$ are known experimentally for
NCO.\cite{Specific_heat, Velocity} The obtained $l$ is compared
with the averaged sample width $W$ (= 0.473 mm), which is taken to
be $2/\sqrt{\pi}$ times the geometrical mean width
$\bar{w}$,\cite{Sun_Comment} and shown in the inset to Fig. 1. It
can be seen that the ratio $l/W$ is only about 0.4 at 0.3 K, which
means that the microscopic scattering on phonons are still
effective at this temperature region.\cite{Berman, Sun_Comment}
Note that if the magnons can contribute to carrying heat at
sub-Kelvin temperatures, the phonon conductivity must be smaller
than the experimental data and the phonon mean free path is even
smaller than the calculated value.

\begin{figure*}
\includegraphics[clip,width=14.5cm]{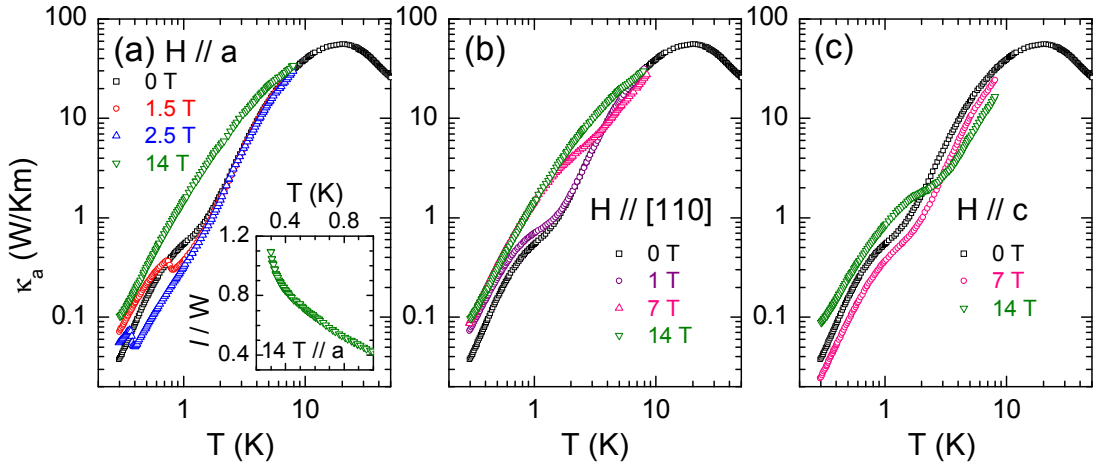}
\caption{(Color online) Temperature dependencies of thermal
conductivity in Nd$_2$CuO$_4$ with magnetic field applied along
the $a$ axis (a), the [110] direction (b) and the $c$ axis (c).
Note that the $\kappa(T)$ curves show step-like transitions at 0.8
and 0.4 K for 1.5 T and 2.5 T fields along the $a$ axis,
respectively. Inset to (a): the ratio of the phonon mean free path
$l$ to the averaged sample width $W$ in 14 T $\parallel a$.}
\end{figure*}

The low-$T$ thermal conductivity of NCO is further studied in the
magnetic fields applied along the $a$ axis, the [110] direction,
and the $c$ axis. For a low magnetic field along the $a$ axis, a
distinct step-like transition shows up in the $\kappa(T)$ curves
below 1.5 K and shifts to lower temperature with increasing field,
as indicated by the 1.5 T and 2.5 T data in Fig. 2(a). This
transition is not observable down to 0.3 K when the magnetic field
is larger than 3 T. It is likely a field-induced AF-ferromagnetic
(spin polarized) transition of Nd$^{3+}$ ions. In the strongest
field (14 T) we applied, the low-$T$ conductivities are increased
and a nearly $T^3$ dependence is presented, which explicitly
indicates that there is strong magnon-phonon scattering in the
zero field and the scattering is almost smeared out in the high
field. This can also be verified from the inset to Fig. 2(a) that
the phonon mean free path in the 14 T field, which is much larger
than that in the zero field, increases with lowering temperature
and approaches the averaged sample width at 0.3 K, demonstrating
that the boundary scattering limit of phonons is nearly
established in the 14 T field and at such low
temperatures.\cite{Zhao_GFO, Sun_Comment} For magnetic field
applied along the [110] direction, a similar result to the
$a$-axis-field case is that the 14 T field can also significantly
enhance the low-$T$ thermal conductivity; in particular, the
thermal conductivity at the lowest temperature is almost the same
as that in 14 T $\parallel a$. However, there seems no
AF-ferromagnetic transition of Nd$^{3+}$ ions for $H
\parallel$ [110], as shown in Fig. 2(b). When the $c$-axis field
is applied, the thermal conductivities display the most complex
dependence on temperature, as shown in Fig. 2(c).

\begin{figure}
\includegraphics[clip,width=8.5cm]{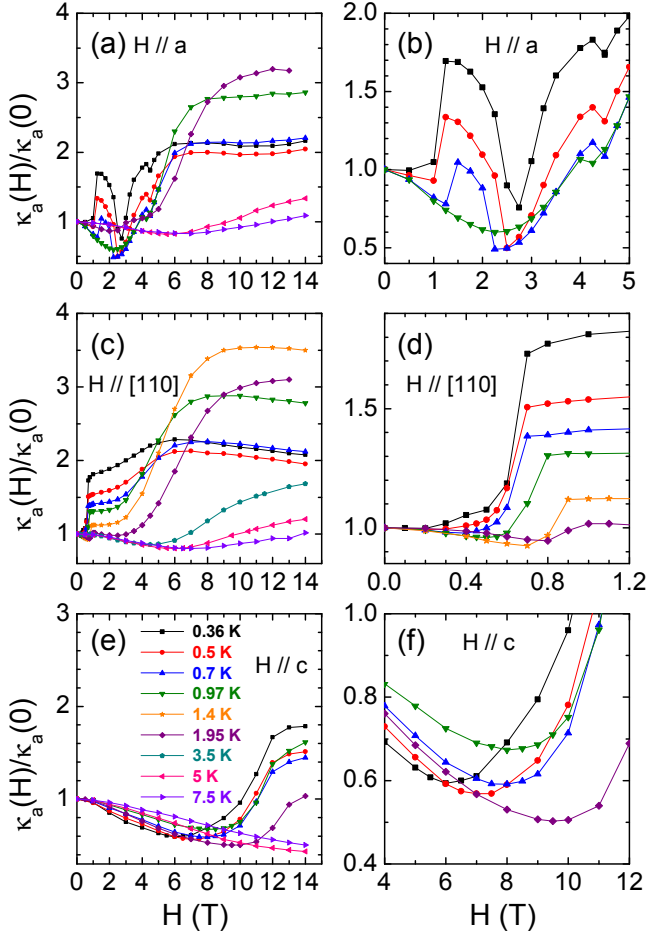}
\caption{(Color online) Magnetic-field dependencies of thermal
conductivity of Nd$_2$CuO$_4$ single crystal with the field
applied along the $a$ axis (a), the [110] direction (c) and the
$c$ axis (e). Panels (b), (d) and (f) zoom in the low-field plots
of panels (a), (c) and (e), respectively.}
\end{figure}

The detailed dependencies of thermal conductivity on the magnetic
field are shown in Fig. 3. One can see from Figs. 3(e) and 3(f)
that as the $c$-axis field increases, $\kappa$ is decreased first
and then increased at high-enough field, which results in a broad
``valley" in the low-$T$ $\kappa(H)$ isotherms. Since both the
Cu$^{2+}$ spin structure and the Nd$^{3+}$ spin structure are not
changeable for $H \parallel c$,\cite{NCO_c-axis} this observation
can only be attributed to the paramagnetic moments. In this
regard, the shifting of the position of the $\kappa(H)$ minimum to
higher field with increasing temperature is indeed compatible with
the typical behaviors of the paramagnetic moments scattering on
phonons.\cite{Sun_PLCO, Sun_GBCO} In NCO, the ``free" spins at
very low temperatures can be either the spin vacancies/defects on
the long-range-ordered Cu$^{2+}$ spin lattice or those on the
ordered Nd$^{3+}$ spin lattice. It is known from the earlier
studies on PLCO and GdBaCo$_2$O$_{5+x}$ that whether the
high-field-limit conductivities are larger than the zero-field
values is determined by the condition whether the magnetic ions
have the zero-field energy splitting.\cite{Sun_PLCO, Sun_GBCO}
Since the low-$T$ $\kappa$ in the high-field limit is apparently
larger than those in zero field, it is likely that the ``free"
spins on the Nd$^{3+}$ sites, whose ground-state doublet can be
split in the zero field,\cite{Structure} rather than the Cu$^{2+}$
free spins,\cite{Sun_PLCO} are responsible for scattering phonons.
In addition, it is clear that the decrease of $\kappa$ in the 14 T
field above 5 K, in contrast to the increase below 2 K, is simply
because the field is not strong enough to remove the paramagnetic
scattering.\cite{Sun_GBCO} Note that the paramagnetic moments
scattering on phonons seems to be much more significant in NCO
than that in PLCO.\cite{Sun_PLCO}

The phonon scattering by paramagnetic moments is known to be
qualitatively isotropic for different field
directions,\cite{Sun_GBCO} so the similar $\kappa(H)$ behaviors to
those for $H \parallel c$ should also contribute to the
$\kappa(H)$ data for $H \parallel ab$. In fact, at relatively high
temperatures above $\sim$ 2 K, the profiles of $\kappa(H)$ for $H
\parallel a$ or [110] could still be understood in the picture of
paramagnetic scattering, whereas at lower temperatures ($T <$ 1 K)
the $\kappa(H)$ isotherms display very different behaviors, which
manifests that some other mechanisms are taking place. For $H
\parallel [110]$, $\kappa$ shows a step-like increase at $\sim$ 0.6 T
and a weak field dependence above this transition field, as shown
in Figs. 3(c) and 3(d). The situation for $H \parallel a$ is
remarkably more complicated. As can be seen from Fig. 3(a), at
sub-Kelvin temperatures, except for a sudden increase at $\sim$ 1
T there is a sharp drop at $\sim$ 2.5 T. One possible mechasim for
these complicated $\kappa(H)$ behaviors is naturally related to
the transitions of spin structures induced by the in-plane field.
It is known that the $a$-axis field and the [110] field cause the
spin-flop transitions of the Cu$^{2+}$ sublattice at $\sim$ 4.5 T
and 0.75 T, respectively.\cite{Cherny, NCO_SF3} Apparently, these
re-orientations have weak effects on the heat transport, since
there is only a very small dip in $\kappa(H)$ curves at 4.5 T for
$H \parallel a$, as shown in Fig. 3(a) and the effect for $H
\parallel$ [110] is not distinguishable from the strong increase
of $\kappa$ at 0.6 T. Therefore, the low-field anomalies of
$\kappa(H)$ for $H \parallel a$ or [110] are most likely caused by
the transitions of the Nd$^{3+}$ sublattice, which have never been
explored in experiments at sub-Kelvin temperatures.

\begin{figure}
\includegraphics[clip,width=8.5cm]{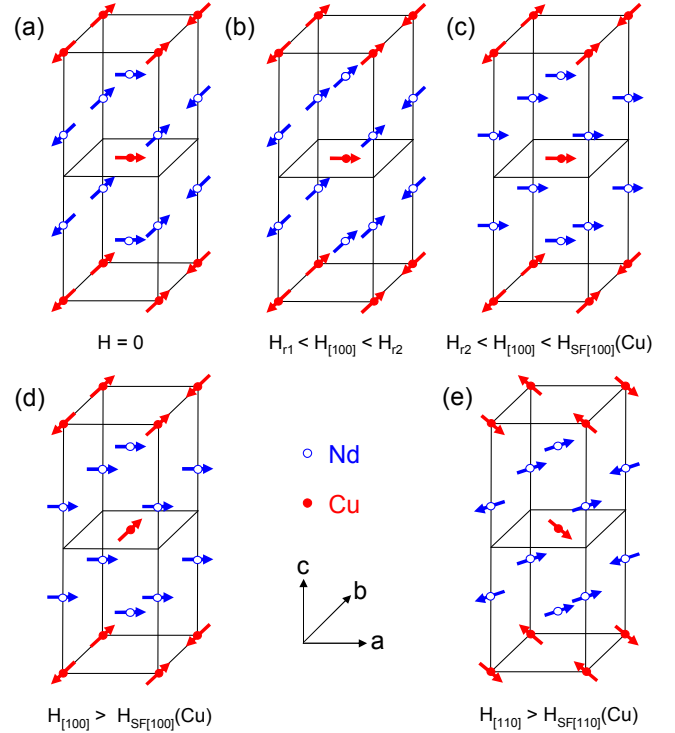}
\caption{(Color online) Very-low-temperature magnetic structures
of Nd$_2$CuO$_4$ in the magnetic fields along two different
in-plane directions. (a) Spin structure in the zero field. (b-d)
Spin structures in the magnetic field along the $a$ axis. $H_{r1}$
and $H_{r2}$ are the critical fields of the spin flop and the spin
polarization transitions of the Nd$^{3+}$ spins, respectively. (e)
Spin structure in the magnetic field along the [110] direction.
$H_{SF[100]}$(Cu) and $H_{SF[110]}$(Cu) are the spin-flop
transition fields of the Cu$^{2+}$ spins for $H \parallel a$ and
[110], respectively.}
\end{figure}

The above $\kappa(H)$ behaviors can suggest the evolution of
Nd$^{3+}$ spin structure (at sub-Kelvin temperatures) for the
in-plane magnetic fields, which are summarized in Fig. 4. In
principle, whether the Nd sublattice rotates together with the
Cu$^{2+}$ spins depends on the competition between
Nd$^{3+}$-Cu$^{2+}$ and Nd$^{3+}$-Nd$^{3+}$ interactions. As
already known, NCO exhibits a noncollinear structure in the zero
field, as shown in Fig. 4(a). With increasing field along the $a$
axis, the Nd$^{3+}$ ions are possible to make spin rotations if
the field is large enough. As shown in Fig. 4(b), when $H > H_{r1}
(\sim$ 1 T) those Nd$^{3+}$ spins pointing along the field
direction could rotate by 90$^{\circ}$ in the $ab$ plane, in other
words, a spin flop transition of Nd$^{3+}$ ions happens. It is
known that the low-energy magnons are usually well populated at
the spin flop transition due to the close of the magnon
gap,\cite{Spin_flop} and meanwhile $\kappa$ shows a sudden
increase at the transition field. It is therefore natural to
conclude that the Nd$^{3+}$ magnons act as heat carriers at very
low temperatures and hence make a positive contribution to the
thermal conductivity. If the magnetic field increases further, the
Nd$^{3+}$ spins gradually turn to the direction of applied field
and are finally polarized at $H = H_{r2}$ ($\sim$ 2.5 T), as
illustrated in Fig. 4(c), which corresponds to a dip in
$\kappa(H)$.\cite{Zhao_GFO, Wang_HMO, Spin_flop} After that the
thermal conductivity starts to recover due to the weakening of
magnon scattering on phonons. At $\sim$ 4.5 T, the Cu$^{2+}$ spin
flop happens and the spin directions are switched to be
perpendicular to the field. This field strength can destroy the
intra-unit Nd$^{3+}$-Cu$^{2+}$ interaction, therefore Nd$^{3+}$
spins can no longer rotate together with Cu$^{2+}$ spins and keep
the alignment along the $a$ axis, as shown in Fig. 4(d). Note that
in the high-field limit, the roles of magnons as either heat
carriers or phonon scatterers are inactive and the plateau of
$\kappa(H)$ also indicates that the paramagnetic scattering on
phonons is smeared out, as the 14 T $\kappa(T)$ data indicate.

In the case of $H \parallel [110]$, there is also a sharp increase
of $\kappa$ at $\sim$ 0.6 T and the magnitude of increase is
almost the same as the increase at 1 T $\parallel a$. This
strongly suggests a magnon heat-transport contribution appearing
at this critical field, which is essentially the same as that for
the Cu$^{2+}$ spin reorientation ($\sim$ 0.75 T in some former
reports).\cite{Cherny, NCO_SF3} Since this field strength is much
smaller than the Nd$^{3+}$-Cu$^{2+}$ interaction, it is expectable
that the Nd$^{3+}$ spins rotate together with the Cu$^{2+}$ spins
and thereby align along the field direction, as shown in Fig.
4(e). Similarly, the low-energy Nd$^{3+}$ magnons are well
populated at this spin flop transition and are able to
significantly transport heat. It is notable that there is no
dip-like feature in higher fields, which indicates that the
Nd$^{3+}$ spin polarization does not happen. This is supported by
the $\kappa(T)$ in Fig. 2(b), where there is no AF-ferromagnetic
transition behavior similar to those in Fig. 2(a).

It is useful to make a quantitative estimation on the Nd$^{3+}$
magnon heat transport from the above data. For example, the
step-like increases of $\kappa$ at 0.36 K is about 0.7 times the
zero-field value, which gives a magnon thermal conductivity of
0.0441 W/Km, for both $H \parallel a$ and $H \parallel$ [110].
Taking a theoretical prediction of the velocity of Nd$^{3+}$
magnons, $\sim$ 10 meV\AA (no experimental observation so
far),\cite{Structure} and assuming the ballistic transport of
magnons at such low temperature,\cite{Li_NCO} we can obtain the
magnon mean free path of about 0.16 mm. This value is quite
reasonable since it is in the same order of magnitude of the
averaged sample width (0.473 mm).

With above understandings on both the paramagnetic scattering and
the magnon heat transport associated with spin-structure
transitions, one can get a complete picture of the $\kappa(H)$
isotherms for the in-plane fields. At sub-Kelvin temperatures, the
sharp increase and the ``dip" at low fields along the $a$ axis are
mainly related to the magnon behaviors, while the high-field
enhancement and the plateau feature are due to the disappearance
of paramagnetic scattering. On the other hand, the step increase
at low fields along [110] mainly results from the magnon heat
transport, and the gradual increase of $\kappa$ at higher field up
to $\sim$ 6 T is likely due to the weakening of paramagnetic
scattering; furthermore, the slow decrease of $\kappa$ at field
above 6 T is related to the suppression of low-energy magnon
excitations in very high fields.

In passing, it is worthy of pointing out that at 1$\sim$2 K the
thermal conductivity is most strongly recovered at very high
in-plane field. It seems to be related to the temperature
dependence of magnon excitations (Nd$^{3+}$); that is, at $T <$ 1
K, the magnon population is negligibly small in zero field because
of the magnon gap, while at 1$\sim$2 K, the critical region of
Nd$^{3+}$ AF transition, the spin fluctuations are significant and
scatter phonons rather strongly. So in this temperature region the
zero-field phonon transport is most strongly damped by not only
the paramagnetic moments but also the magnetic excitations and it
can be remarkably recovered by applying high in-plane field.

It is intriguing to compare the present work with the earlier
ones. It has been concluded from the very-low-$T$ thermal
conductivity in the zero field and strong in-plane field, which is
much higher than the Cu$^{2+}$ spin-flop transition, that the
high-field-induced enhancement of $\kappa$ is a direct
contribution from Nd$^{3+}$ magnon heat transport.\cite{Li_NCO}
The present data essentially support the capability of Nd$^{3+}$
magnon transporting heat, but only in relatively low fields. Base
on the detailed field dependence of $\kappa$, particularly for $H
\parallel c$, it is able to be clarified that the enhancement of
$\kappa$ at very high fields is mainly due to the weakening of
paramagnetic scattering on phonons. The reason that the earlier
work did not notice the importance of paramagnetic scattering is
that the $\kappa(T)$ data for $H \parallel c$ were taken only at
10 T for the comparison with the in-plane-field data.\cite{Li_NCO}
It is a coincidence that in this field, the thermal conductivities
in the whole temperature range are not likely to be larger than
the zero-field values, as shown in Fig. 3(e). Therefore, the
present data provide a very important supplement to the earlier
works and lead to a more accurate conclusion.

\section{Conclusions}

In summary, we study the heat transport of a parent insulator of
high-$T_c$ superconductors Nd$_2$CuO$_4$ at low temperatures down
to 0.3 K and in magnetic fields up to 14 T. It is found that in
zero field the low-$T$ thermal conductivity is purely phononic
with rather strong scatterings from paramagnetic moments and
Nd$^{3+}$ magnetic excitations. In high magnetic field along
either the $c$ axis or the $ab$ plane, the low-$T$ thermal
conductivity can be significantly enhanced because of the
weakening of magnetic scattering. An interesting finding is the
drastic changes of $\kappa$ at low fields along the $a$ axis or
the [110] direction, which demonstrates the field-induced spin
flop or spin polarization of Nd$^{3+}$ spin lattice. At sub-Kelvin
temperatures, the Nd$^{3+}$ magnons can act as heat carriers in
the spin-flopped state, however, their transport can exist only in
some intermediate field regime and is suppressed by an succeeding
spin-polarization transition for $H \parallel a$. In the field
along the [110] direction, the magnon transport can be active in a
broader field region but is also weakened in very high field. An
evolution of the magnetic structure with the in-plane field can be
suggested based on these field dependencies of $\kappa$. One may
note that although the magnons can hardly affect the phonon heat
transport for those superconducting samples, because of the
disappearance of the long-range AF order, the paramagnetic moments
can still effectively scatter phonons.

\begin{acknowledgements}

This work was supported by the Chinese Academy of Sciences, the
National Natural Science Foundation of China, the National Basic
Research Program of China (Grants No. 2011CBA00111 and No.
2009CB929502), and the RFDP (Grant No. 20070358076).

\end{acknowledgements}

\end{document}